\documentclass[a4paper]{jpconf}
\usepackage[utf8]{inputenc} 
\usepackage{graphicx}
\usepackage{amsmath}
\usepackage{cite}

\DeclareMathOperator{\im}{Im}

\begin{document}
\title{In-medium Spectral Functions in a Coarse-Graining Approach}
\author{Stephan Endres$^{1, 2}$, Hendrik van Hees$^{1, 2}$, Janus Weil$^{1, 2}$ and Marcus Bleicher$^{1, 2}$}
\address{$^{1}$Institut f\"{u}r Theoretische Physik, Universit\"{a}t
  Frankfurt, Max-von-Laue-Stra{\ss}e 1, 60438 Frankfurt, Germany} 
\address{$^{2}$Frankfurt Institute for Advanced Studies,
  Ruth-Moufang-Stra{\ss}e 1, 60438 Frankfurt, Germany} 
\ead{endres@th.physik.uni-frankfurt.de}
%%%%%%%%%%%%%%%%%%%%%%%%%%%%%%%%%%%%%%%%%%%%%%%%%%%%%%%%%%%%%%%%%%%%%%%%%%%%
\begin{abstract}
  We use a coarse-graining approach to extract local thermodynamic
  properties from simulations with a microscopic transport model by
  averaging over a large ensemble of events. Setting up a grid of small
  space-time cells and going into each cell's rest frame allows to
  determine baryon and energy density. With help of an equation of state
  we get the corresponding temperature $T$ and baryon-chemical potential
  $\mu_{\mathrm{B}}$. These results are used for the calculation of the thermal
  dilepton yield. We apply and compare two different spectral functions
  for the $\rho$ meson, firstly a calculation from hadronic many-body
  theory and secondly a calculation from experimental scattering
  amplitudes. The results obtained with our approach are compared to
  measurements of the NA60 Collaboration. A relatively good description 
  of the data is achieved with both spectral functions. However, the hadronic
  many-body calculation is found to be closer to the experimental data
  with regard to the in-medium broadening of the spectral shape.
\end{abstract}
%%%%%%%%%%%%%%%%%%%%%%%%%%%%%%%%%%%%%%%%%%%%%%%%%%%%%%%%%%%%%%%%%%%%%%%%%%%%
\section{Introduction}

Although it is widely accepted nowadays that the dynamics of strong
interactions are governed by quantum chromodynamics (QCD), we still lack
a full understanding of the QCD phase structure. At low energies and
densities, the relevant degrees of freedom are hadrons, i.e., composite
objects of quarks and gluons. However, for sufficiently large energies
and/or densities the creation of a deconfined phase with free quarks and
gluons is expected. Also the symmetry pattern of QCD is assumed to
change - e.g., chiral symmetry which is broken in the vacuum but
expected to be restored at some finite temperature. In consequence, 
the study of the in-medium properties of hadrons plays an important role 
especially for a better understanding of the non-perturbative part of QCD
\cite{Leupold:2009kz, Rapp:1999ej}. Experimentally, heavy-ion collisions 
at ultra-relativistic energies are a good possibility to create hot and
dense matter in the laboratory. However, the measurement of in-medium
modifications is difficult, as the fireball has a lifetime of only a few
fm/$c$, and all the hadronic observables only give information on the
final freeze-out. Here dileptons offer the big advantage that they do
not interact strongly and can leave the hot and dense fireball unscathed
after their production. But this has a drawback, too. As lepton pairs
are produced via many different processes at all stages of the reaction,
their spectra reflect the whole evolution of the collision. What one can
measure are only time integrated spectra. This is a challenge for
theory, as it requires a reliable description of the entire reaction
dynamics.

In these proceedings a coarse-graining approach \cite{Endres:2014zua} is 
used (following a similar ansatz as proposed by Huovinen et al.\,\cite{Huovinen:2002im}), 
which allows to combine in-medium spectral functions with an underlying microscopic
description of the reaction dynamics. We argue that this is a good and
realistic compromise between transport approaches, where an inclusion of
medium modifications of spectral functions is difficult, and fireball
models that use a simplified description of the reaction evolution but
where the application of spectral functions is straightforward. In
contrast to common hydrodynamic models, its advantage is the unified
description of the whole reaction dynamics, from the first hadron-hadron
collisions to the final freeze-out (except for a different treatment of 
the very low temperature cells, see section \ref{model}).

%%%%%%%%%%%%%%%%%%%%%%%%%%%%%%%%%%%%%%%%%%%%%%%%%%%%%%%%%%%%%%%%%%%%%%%%%%%%%
\section{\label{model} The Model}

The coarse-graining is based on the time-step-wise output from
simulations with the Ultra-relativistic Quantum Molecular Dynamics model
(UrQMD). It is a non-equilibrium transport approach and describes a
nuclear reaction in terms of classically propagated hadrons combined
with subsequent elastic and inelastic binary scatterings
\cite{Bass:1998ca,Bleicher:1999xi,Petersen:2008kb}. 
To extract thermodynamic properties, we take an ensemble of UrQMD events
($\approx 1000$) and average over them, such that we get a relatively
smooth phase-space distribution function. Hereby the UrQMD output is set
on a space-time grid with $\Delta t= 0.2 \; \mathrm{fm}/c$ and
$\Delta x = 0.8 \;\mathrm{fm}$. The choice of the cell size is a compromise 
between the highest possible resolution and the memory and computational requirements. 
However, we checked that a smaller size does not improve or change
the results. 

In each cell the baryon four-flow is then calculated and 
Eckart's definition \cite{Eckart:1940te} is used to perform a Lorentz
boost into the local rest frame of the cell. This enables to
calculate local baryon and energy densities. To extract the temperature
$T$ and baryon-chemical potential $\mu_{\mathrm{B}}$ we use the equation
of state of a hadron resonance gas \cite{Zschiesche:2002zr} that
includes the same hadronic degrees of freedom as the UrQMD model. To
account also for dilepton emission from the quark-gluon plasma phase, we
use a second equation of state for temperatures above 170\;MeV that has
been fitted to Lattice calculations \cite{He:2011zx}. As the values of
$T$ of the two EoS agree almost perfectly in the temperature range from
150 to 170\;MeV, a smooth transition is guaranteed. Note, however, that
the chemical potential is set to zero for $T> 170\;\mathrm{MeV}$, as the
Lattice fit is only done for vanishing quark chemical potential. In 
consequence we can not have a smooth transition for $\mu_{B}$. 
(Besides $T$ and $\mu_{\mathrm{B}}$, we also calculate the effective
baryon density, $\rho_{\mathrm{eff}}$, in each cell, as it is the input
for one of the spectral functions implemented here, see section 3
below). When we know the thermodynamic properties of the cell, the
dilepton emission per four-volume and four-momentum is calculated
according to \cite{Rapp:2013nxa}
\begin{equation}
\label{rate}
\frac{\mathrm{d}N_{ll}}{\mathrm{d}^{4}x\mathrm{d}^{4}p} =
-\frac{\alpha_{\mathrm{em}}^{2}}{\pi^{3}M^{2}} L(M) f^{\mathrm{B}}(E;T)\im
\Pi^{(\mathrm{ret})}_{\mathrm{em}}(E,p;\mu_{\mathrm{B}},T) 
\end{equation}
with the dilepton phase-space function $L$ and the Bose-distribution
function $f^{\mathrm{B}}$. In case of the thermal $\rho$ emission the
retarded electromagnetic current-current correlator is given by the
in-medium propagator of the meson,
$\im\Pi_{\mathrm{em}}=(m_{\rho}^{2}/g_{\rho})^{2}\im D_{\rho}$ (where
$m_{\rho}$ respectively $g_{\rho}$ denote the bare mass and the coupling
strength of the $\rho$).

Note that the original dilepton emission rate, as derived by McLerran
and Toimela \cite{McLerran:1984ay}, assumes chemical equilibrium.
However, here we use a generalization for an off-equilibrium situation
with finite pion chemical potentials. In this case an additional squared
fugacity factor $z^{2}_{\pi}=\exp\left(2\mu_{\pi}/T\right)$ enters in
equation (\ref{rate}) for the thermal $\rho$ emission. Within our
approach the pion chemical potential is extracted in each cell in the
relativistic Boltzmann approximation. A positive value for $\mu_{\pi}$
means that the corresponding number of produced pions is larger than one
would obtain in complete chemical equilibrium
\cite{Bandyopadhyay:1993qj}. This is what we also find when using the
input from the UrQMD model. As even small values of $\mu_{\pi}$ give
a strong increase on the dilepton yield
\cite{Koch:1992rs}, it is important to include this effect in the
calculations.

Although the focus is on the in-medium effects of the $\rho$ in the
present study, we need to consider also the other sources that
contribute to the dilepton excess yield. For a full description one has
to calculate the emission from the QGP, i.e., in our case for
temperatures above 170\;MeV. For this we use the rates from lattice
calculations, which are extrapolated for finite momenta
\cite{Ding:2010ga}. As we deal with extremely high densities of hadronic
matter in the heavy-ion collisions under consideration, multi-pion
interactions will also give a significant contribution for the dilepton
yield. The rates used in our approach take vector-axialvector mixing
into account \cite{vanHees:2006ng, vanHees:2007th}.

Finally, we have to consider the contribution from those cells, where
the density (respectively temperature) is very low and therefore the
assumption of a thermal emission is not reasonable. For these cells we
simply take the $\rho$ mesons from the UrQMD calculation directly and
apply a shining procedure \cite{Li:1996mi, Schmidt:2008hm}. As this is
only done for the cells in which no thermal emission is considered, we
avoid double counting.

For more details on the coarse-graining approach, the reader is referred to 
reference \cite{Endres:2014zua}.
%%%%%%%%%%%%%%%%%%%%%%%%%%%%%%%%%%%%%%%%%%%%%%%%%%%%%%%%%%%%%%%%%%%%%%%%%%%%%

\section{\label{SF} Spectral Functions}

Only few approaches exist which model the influence of both finite 
temperature and baryon density on the $\rho$ spectral function. We here apply 
and compare two widely used spectral functions: On the one hand, the 
hadronic many-body calculation by Rapp and Wambach (Rapp SF)
\cite{Rapp:1999us} which has previously proven to be a good description
of experimental results
\cite{vanHees:2006ng,vanHees:2007th,Rapp:2013nxa}. And secondly the more 
model-independent approach by Eletsky and others (Eletsky SF)
\cite{Eletsky:2001bb} for which the self-energies of the $\rho$ are
calculated using empirical scattering amplitudes from resonance
dominance. In the Rapp SF the propagator of the $\rho$ takes the form 
\begin{equation}
  D_{\rho}(E,p)=\frac{1}{M^{2}-M_{\rho}^{2}-\Sigma_{\rho\pi\pi}(M)-\Sigma_{\rho
      M}(E,p)-\Sigma_{\rho B}(E,p)}, 
\end{equation} 
which includes interactions of the pion cloud with the $\rho$
($\Sigma_{\rho\pi\pi}$) and direct scatterings off mesons and baryons
($\Sigma_{\rho M}$, $\Sigma_{\rho B}$). The meson-gas effects include
interactions with the most abundant $\pi$, $K$ and $\rho$ mesons and are
saturated with resonances up to 1.3\;GeV \cite{Rapp:1999us,Rapp:1999qu}.
Similarly, the effects of the baryonic matter enter via direct
$\rho N \rightarrow B$ scattering and resonance hole excitations
\cite{Rapp:1997ei,Urban:1998eg}. For the Eletsky SF the $\rho$-meson 
self-energy is calculated within a heat bath of nucleons and pions.
The in-medium modifications are then consequently determined 
by $\rho N$ and $\rho\pi$ scatterings. In the low-energy 
regime the calculation of the scattering amplitudes
$f_{\rho a}(s)$ includes the resonances from the partial wave analysis
by Manley \cite{Manley:1992yb}, while at higher energies a Regge
parametrization is applied. The contribution to the self-energy from
the scattering of a $\rho$ meson with a hadron is calculated in
low-density expansion as
\begin{equation} \Sigma_{\rho a} (E,p) = -4\pi \int
\frac{\mathrm{d}^{3}k}{(2\pi)^{3}}
n_{a}(\omega)\frac{\sqrt{s}}{\omega}f_{\rho a}(s) 
\end{equation} 
with $\omega^{2}=m_{a}^{2}+k^{2}$ and the occupation number
$n_{a}$. Note that in this model the scattering amplitudes are evaluated
on the mass shell and therefore only depend on $p$ here.

When comparing the two approaches for the spectral function, there are
some caveats one has to consider. Especially it is important to know
that the implementation of baryonic effects is not the same. In the
parametrized version of the Rapp SF \cite{RappSF}, which we use for our
calculations, the baryonic effects enter via an effective baryon density
$ \rho_{\text{eff}}=\rho_{\mathrm N} + \rho_{\bar{\mathrm N}} + 0.5
\left(\rho_{ \mathrm B^{*}} + \rho_{\bar{\mathrm B}^{*}} \right)$.
Firstly, this takes into account that the interaction of the $\rho$
meson is the same with baryons and anti-baryons.  And secondly it
reflects the fact, that the scattering with excited baryons has less
effect on the spectral shape \cite{vanHees:2007th}. As the Eletsky SF is
determined for a gas of $\pi$ and $N$ only, not considering the effects
of anti-baryons and excited resonance states, the baryon effects enter
via a nucleon chemical potential $\mu_{N}$. In the non-interacting
hadron gas EoS, which we apply when determining $T$ and $\mu_{\mathrm{B}}$,
this $\mu_{N}$ is identical with the baryon-chemical potential
$\mu_{\mathrm{B}}$. However, it is important to bear in mind that the
degrees of freedom are different for the underlying UrQMD model and the
hadron gas EoS on the one side and the $\pi N$ gas, for which the
scattering amplitudes of the Eletsky SF were calculated, on the other
side. At SPS energies, a significant fraction of the baryons is found in
an excited state in the UrQMD calculations, in conflict with the
assumption of a gas of pions and nucleons only. Here a significantly
higher degree of consistency is given with the hadron-resonance gas used
for the hadronic many-body calculations by Rapp.
%%%%%%%%%%%%%%%%%%%%%%%%%%%%%%%%%%%%%%%%%%%%%%%%%%%%%%%%%%%%%%%%%%%%%%%%%%%%

\section{\label{results} Results}
\begin{figure}
\includegraphics[width=.52\textwidth]{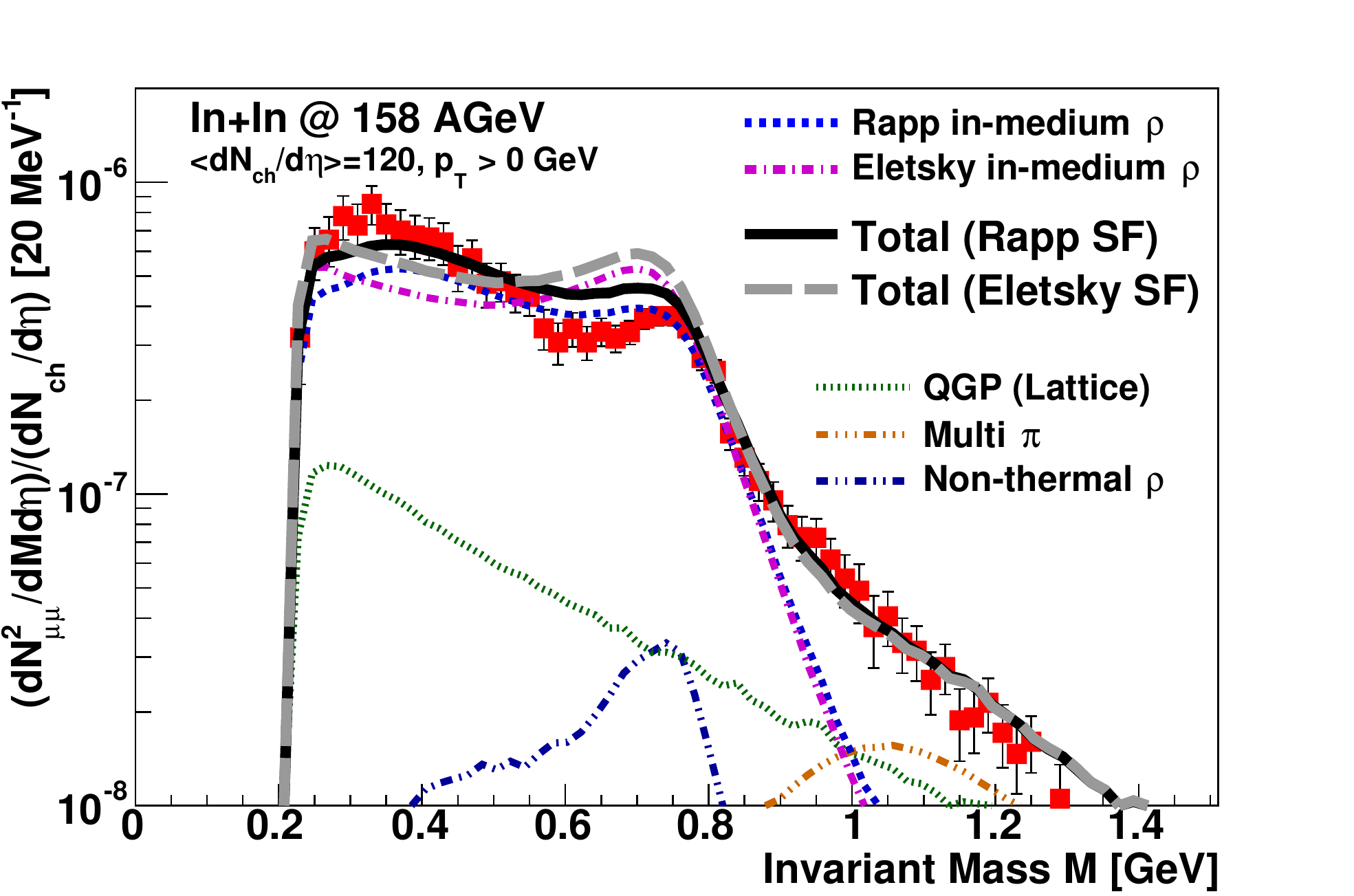}
\includegraphics[width=.52\textwidth]{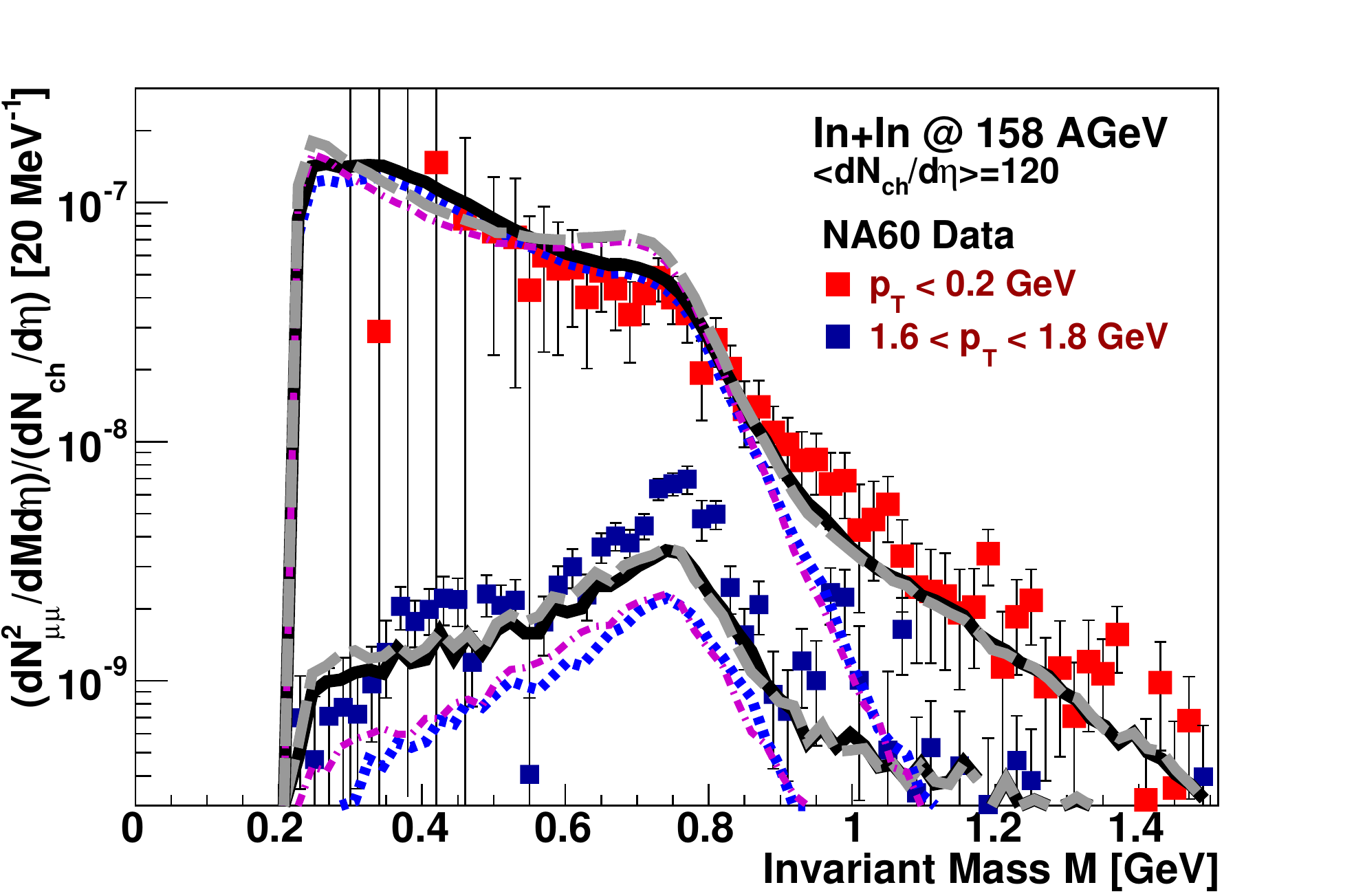}
\caption{(Color online) Invariant mass spectra of the dilepton excess
  for In+In collisions at $E_{\mathrm{lab}}=158 \,A \mathrm{GeV}$ for
  the low-mass region for all $p_{t}$ (left plot) and two distinct
  transverse momentum bins (right plot). The results are compared to the
  experimental data from the NA60 Collaboration \cite{Arnaldi:2008er,
    Specht:2010xu, Arnaldi:2008fw}.}
\label{fig1}
\end{figure}
The resulting invariant mass spectra from our coarse-graining
calculation are presented in Figure \ref{fig1}. The left plot shows the
total $p_{t}$-integrated yield. In general we achieve a rather good
description of the measured spectra \cite{Arnaldi:2008er, Specht:2010xu, Arnaldi:2008fw} 
with both spectral functions. A somewhat stronger melting of the peak
at the $\rho$ pole mass is however visible in case of the Rapp SF. 
It only slightly overshoots the NA60 data by about 20\% in the 
mass range from 0.6 to 0.8\;GeV, while the Eletsky SF
gives a yield that is almost a factor of two above the data. This
finding is in line with the fact that the Eletsky approach resembles a
virial expansion, that is valid only for low densities. The broadening
is mainly due to the presence of baryonic matter which in such heavy-ion
collisions can easily reach several times normal matter
density. Consequently it is clear that the low-density approximation
will underestimate the effect. On the contrary, we get more dilepton
yield in the mass range from 0.3 to 0.5\;GeV with the Rapp SF, in better
agreement with the data than the Eletsky SF. The latter additional
enhancement was partly attributed to the inclusion of off-shell $\rho$-$N$ 
resonances \cite{Rapp:1999ej}, whereas in the Eletsky SF the scattering 
amplitudes are calculated on-shell only. The overall shape of the
invariant-mass spectrum is clearly better described when applying the
Rapp SF; we find a stronger broadening here, in good agreement with the
data. The other thermal dilepton sources, i.e., the quark-gluon plasma
and the multi-pion interactions, only give relatively small
contributions at low masses. However, they become significant in the mass
range above 1\;GeV, which is not considered in detail here. A relatively
small contribution stems also from the non-thermal $\rho$.

In the right plot of Figure \ref{fig1} we see the results for two
transverse-momentum bins. The upper curves (with red data points) denote
the results for $p_{t} < 0.2\; \mathrm{GeV}$, while the lower results
(blue data points) are for momenta between 1.6 and 1.8 GeV. One can see
from the comparison, that the most dominant modification of the spectral
functions is observed for low momenta. Here - as in the
$p_{t}$-integrated spectrum - we find a stronger broadening for the Rapp
SF, better describing the data around the pole mass. For the higher
momentum window, the spectral shape of the $\rho$ resembles more its
vacuum shape. Consequently, the differences between the spectral
functions become less dominant. However, for high $p_{t}$  there remains 
some excess of the data above the model results around the $\rho$ pole mass.
Obviously we miss some of the "freeze-out" $\rho$ contribution here.
Note also that the thermal $\rho$ contribution clearly dominates for 
invariant masses up to 1\;GeV at low transverse momenta, while for the 
higher $p_{t}$ bin it makes up only roughly 50\% in this mass region, 
i.e., the QGP contribution and the non-thermal $\rho$ become more relevant 
here (not shown in the plot for reasons of lucidity).
%%%%%%%%%%%%%%%%%%%%%%%%%%%%%%%%%%%%%%%%%%%%%%%%%%%%%%%%%%%%%%%%%%%%%%%%%%%%%
\section{\label{summary} Summary \& Conclusions}

In general, the coarse-graining approach has proven its applicability
for the description of experimental dilepton yields. Although the two
spectral functions used in the present work stem from two rather different
calculations, both give a reasonable description of the experimentally 
measured dilepton yields at SPS energy within the coarse-graining 
approach. In comparison, however, the description of the broadening (which
is mainly due to scattering of the $\rho$ with baryons) is more
realistic in the hadronic many-body calculation by Rapp and Wambach.
%%%%%%%%%%%%%%%%%%%%%%%%%%%%%%%%%%%%%%%%%%%%%%%%%%%%%%%%%%%%%%%%%%%%%%%%%%%%%

\ack

We thank Ralf Rapp for providing the parametrization of the spectral
function and many fruitful discussions. This work was supported by BMBF,
HIC for FAIR and H-QM.

\section*{References}
\bibliography{Proc_Endres_FAIRNESS2014}% Produces the bibliography via BibTeX.

\end{document}